%% file: paper.tex
\documentclass[letterpaper,twocolumn,10pt]{article}
\usepackage{usenix2019_v3}

\usepackage{graphicx}

\usepackage[english]{babel}
\usepackage{blindtext}

\usepackage{xspace}
\newcommand{\fitnets}{\textsf{FitNets}\xspace}

\usepackage{amsmath}
\usepackage{xfrac}  

\usepackage{tabu}
\usepackage{multirow}
\usepackage{booktabs}  

\usepackage{subcaption}
\usepackage{float}

\newcommand{\est}{\ensuremath{f_n}\xspace}
\newcommand{\real}{\ensuremath{f}\xspace}
\newcommand{\qs}[2]{\ensuremath{\mathrm{QS}({#1},\,{#2})}\xspace}
\newcommand{\s}[2]{\ensuremath{\mathrm{S}({#1},\,{#2})}\xspace}

\newcommand{\remove}[1]{}

\begin{document}

\date{}

\title{\fontsize{16.15}{21} \bf FitNets: An Adaptive Framework to Learn Accurate Traffic Distributions}

\author{
  {\rm Alexander Dietm\"uller}\\
  ETH Z\"urich
  \and
  {\rm Albert Gran Alcoz}\\
  ETH Z\"urich
  \and
  {\rm Laurent Vanbever}\\
  ETH Z\"urich
}

\maketitle

\begin{abstract}
  Learning precise distributions of traffic features (e.g., burst sizes, packet
  inter-arrival time) is still a largely unsolved problem despite being critical
  for management tasks such as capacity planning or anomaly detection. A key
  limitation nowadays is the lack of feedback between the control plane and the
  data plane. Programmable data planes offer the
  opportunity to create systems that let data- and control plane to work
  together, compensating their respective shortcomings.

  We present \fitnets, an adaptive network monitoring system leveraging
  feedback between the data- and the control plane to learn accurate traffic
  distributions. In the control plane, \fitnets relies on Kernel Density Estimators
  which allow to provably learn distributions of any shape. In the data plane,
  \fitnets tests the accuracy of the learned distributions while dynamically
  adapting data collection to the observed distribution fitness, prioritizing under-fitted features.

  We have implemented \fitnets in Python and P4 (including on commercially available programmable switches) and tested it on real and synthetic traffic traces. \fitnets is practical: it is able to estimate hundreds of distributions from up to 60 millions samples per second, while providing accurate error estimates and adapting to complex traffic patterns.
\end{abstract}

\input{section_introduction}
\input{section_motivation}
\input{section_overview}
\input{section_design}
\input{section_evaluation}
\input{section_related}
\input{section_conclusion}

\newpage
\appendix
\input{appendix}

\bibliographystyle{plain}
\bibliography{generated_refs.bib}

\end{document}

%% file: section_introduction.tex
\section{Introduction}
\label{sec:introduction}

Collecting reliable traffic statistics is one of the most fundamental problems
of network monitoring and is crucial for successfully operating a network.
Traffic statistics (i.e., traffic distributions) are indeed used in a wide
range of network management tasks including capacity planning, traffic
engineering, billing, and fault-detection.

Despite its importance, collecting such statistics remains a challenge, in
particular in Internet Service Providers (ISPs). As ISPs do not have control
over the end hosts, operators have no choice but to collect traffic statistics
using suboptimal in-network monitoring tools. Typical example of such
monitoring tools include NetFlow~\cite{claise_cisco_2004} and
sFlow~\cite{phaal_inmon_2001}. As both such tools are based on random packet
sampling, their outputs typically only cover a very small fraction of the
network traffic, and quite poorly so. Studies have shown that simple packet sampling strategies
(e.g., 1 in $N$) fail to estimate key statistics such as the
distribution of flow sizes~\cite{tune2008towards,
  duffield2003estimating}.

With programmable networking devices using languages such as
P4~\cite{bosshart_p4:_2014}, new approaches to network monitoring have emerged
in an attempt to cast off the drawbacks of packet
sampling by getting ``closer to the information'', moving processing to the data plane.
The two most prominent advances in this direction are: (i) processing
\emph{queries} by executing operations such as \texttt{filter} and \texttt{map}
on all the traffic, returning only the result to the control
plane~\cite{gupta_sonata:_2017,narayana_language-directed_2017};
and (ii) aggregating frequency statistics using \emph{sketches}, probabilistic
data structures from which the control-plane can extract information
approximately (e.g. e.g.~\cite{yu_software_2013,liu_one_2016}).
However, both approaches are fundamentally limited by the capabilities of
programmable data-planes, which support only simple instructions and limited
memory access in order to process traffic at line rate.

To sum up, a gap remains between sampling-based solutions, in which a
powerful control plane acts on sparse information, and sampling-free solutions,
which are limited by the capabilities of programmable data-planes.

\begin{figure}[h]
  \centering
  \includegraphics[width=0.8\columnwidth]{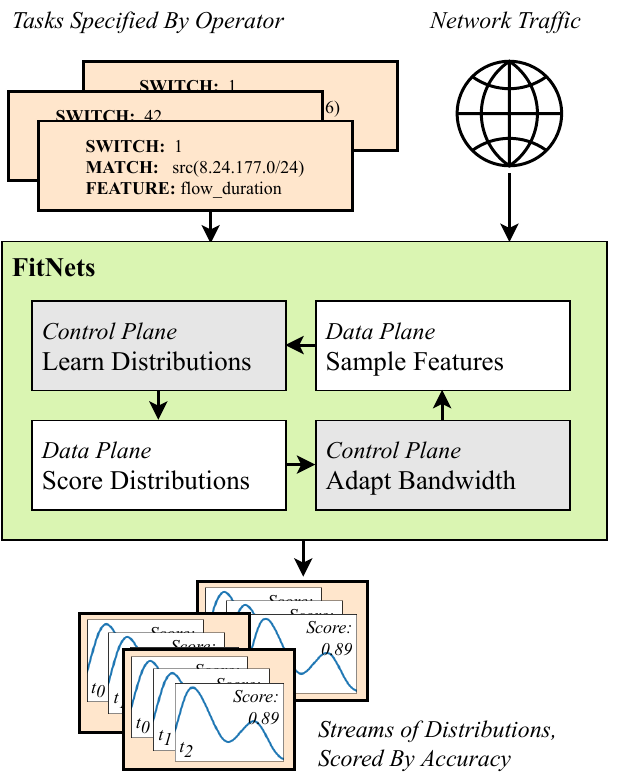}
  \caption{\fitnets learns from traffic to optimizes monitoring.}
  \label{fig:firstpage_overview}
\end{figure}

\medspace
\paragraph{\fitnets}
This paper presents \fitnets, an adaptive approach and a system which bridges this
gap. \fitnets combines monitoring in both data- and control plane through
feedback. More specifically, \fitnets uses the computational power of the control
plane to estimate distributions, and the line rate processing capabilities of
the data plane to to evaluate the accuracy of the learned distributions.
Intuitively, this division represents the fact that learning is
complex (i.e., has to be done in software), but verifying it is easy (i.e.,
can be done in hardware).

Similarly to \emph{active learning}, \fitnets control plane optimizes sampling to
collect information where it is needed most to improve the estimates. This
flexibility in adapting the sampling rate enables \fitnets to meet a wide range of
management objectives. Among others, it can \emph{minimize} the required
sampling rate while maintaining fixed accuracy requirements, and on the other
hand, \fitnets can \emph{maximize} the accuracy (in a max-min approach) for fixed
sampling rates. We want to note that \fitnets is not limited to these two
approaches and can easily by extended to arbitrary adaptation strategies.

We have fully implemented \fitnets including its control plane (in Python) and its
data plane components (in P4, including with support for commercially available
programmable switches). Our evaluation using real and synthetic traffic traces
shows that \fitnets is practical: It is able to estimate hundreds of distributions
from up to 60 millions samples per second, while providing reliable accuracy
estimates and adapting to complex traffic patterns.


%% file: section_motivation.tex
\section{Overview}
\label{sec:motivation}

In this section, we provide an overview of \fitnets using a motivating
example. \fitnets' pipeline consists of four main connected steps (see
Figure~\ref{fig:firstpage_overview}): (i) estimating distributions from
samples (control plane); (ii) scoring estimates on \emph{all traffic}
(data plane); (iii) normalizing the the scores to reliably estimate the
current accuracy (control plane); and (iv) adapting the sampling to
operator goals (control plane).

\paragraph{Example} Alice and Bob are operators responsible for network monitoring
of two internet service providers, $A$, and $B$.
Both have a set of \emph{monitoring tasks}: They have defined
(i) where they want to collect data, i.e. sample traffic; (ii) which parts of
traffic they are interested in, e.g. certain prefixes; and
(iii) which statistics they are interested in, e.g. the duration of
traffic bursts, or the time packets spend in queues.
While their tasks are similar, their demands are different:
\begin{itemize}
      \item $A$ wants to ensure smooth operation while minimizing cost.
            The monitoring tasks need to meet a certain level of accuracy, while
            requiring as little sampling as possible.
      \item $B$ wants to invest in network research and is consequently
            interested in high accuracy across all tasks. However, the network needs
            remain operational, and there is a hard limit on the available bandwidth.
\end{itemize}
Despite their different goals, both $A$ and $B$ ultimately try to determine the
optimal sampling rate across different tasks.

A naive monitoring system that returns only the raw sampled data lacks
information on the accuracy of statistics that can be extracted. Typically,
only
a few percent of network traffic can be realistically monitored, and attempting
to
estimate the accuracy based on only this fraction of traffic potentially misses
the majority of information, e.g. over-estimating how well a statistic
represents traffic, because it fits the small sample well. To make matters
worse, network statistics typically both vary greatly in their complexity as
well as may change over time, which makes makes manual analysis infeasible, as
by the time the operator has evaluated the estimate, the underlying traffic
has likely changed already.

Consequently, both Alice and Bob end up either sampling too much or too
little traffic for the respective statistics. $A$ resorts to over-provisioning
to ensure that the minimum accuracy requirements are met, resulting in
unnecessary cost, while $B$ tries to manually adjust the sampling for each task,
and is prone to miss new trends or react slowly..

\fitnets provides an adaptive framework addressing both Alice's and Bob's
problems. It is based on two core concepts:
\begin{description}
      \item[Learning] \fitnets does not only return raw data to the operator.
            Instead, it deeply integrates estimating the required statistics from the
            data, and uses programmable data planes to estimate the accuracy of the
            learned statistics on all network traffic.

      \item[Adaptation] \fitnets offers programmable adaptation. It can be
            programmed to automatically adapt sampling to meet operator-specified goals.
\end{description}

Concretely, $A$ and $B$ first specify their respective monitoring tasks
using a simple query language (which we introduce in
section~\ref{sec:implementation}) and configure the adaptation objective. $A$'s
objective corresponds to minimizing the sampling rate for given accuracy
requirements, while $B$'s objective corresponds to maximizing the accuracy for
a given bandwidth. Both objectives are supported by \fitnets.
Consequently, \fitnets enables monitoring that adapts to operator-specified
requirements in the face of complex and changing network statistics.

%% file: section_overview.tex
\section{Estimation of Traffic Distributions}
\label{sec:overview}

Estimating traffic distributions and providing both reliable as well as
comparable accuracy
scores are the core concepts that allow \fitnets to produce a stream of
probability distributions that can be used effectively, e.g., to optimize
sampling or to enable operators to make informed decisions based on the
observed accuracy of estimation.
In summary, \fitnets' estimation goes through three steps:
\begin{description}
    \item[Estimation] The distributions of features are estimated
          from samples using \emph{Kernel Density Estimation}
          (section~\ref{ssec:overview_kde}).
    \item[Scoring] The estimates are scored using \emph{proper
              scoring rules}, which can be computed in the data plane
          (section~\ref{ssec:overview_scoring}).
    \item[Normalization] The accuracy of estimates is computed by normalizing
          the score, which requires
          \fitnets to estimate the (unknown) optimal possible score
          (section~\ref{ssec:overview_score_estimation}).
\end{description}
This allows \fitnets to return distributions for arbitrary traffic features, along
with normalized scores, which are comparable even for different distributions.
Additionally, the low complexity of score computation enables data-plane
scoring.

\subsection{Kernel Density Estimation}
\label{ssec:overview_kde}

Estimating distributions for network traffic features is challenging, as the
underlying distributions can have complex shapes (e.g., assumptions such as
normality do not necessarily hold) and may change over time.
Furthermore, to cope with the large volume and dynamic nature of
network traffic, the estimates must be both efficient and fast to compute.

In this setting,
Kernel Density Estimators (KDEs) are useful non-parametic estimators,
which do not assume any shape of the underlying distribution
and can be efficiently computed using
the fast Fourier transformation (FFT)~\cite[p.~183]{wand_kernel_1995}.
Additionally, the mean integrated square error (MISE) of a KDE can be
asymptotically approximated for a given training sample size, even if the true
distribution is unknown.

\begin{figure}
    \begin{subfigure}{0.48\linewidth}
        \centering
        \includegraphics[width=\linewidth]{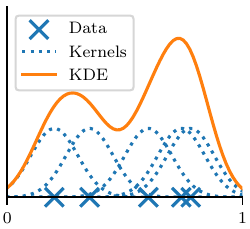}
        \subcaption{The KDE is a sum of kernels centered at samples.}
        \label{fig:overview_kde_illustration}
    \end{subfigure}\hfill
    \begin{subfigure}{0.48\linewidth}
        \centering
        \includegraphics[width=\linewidth]{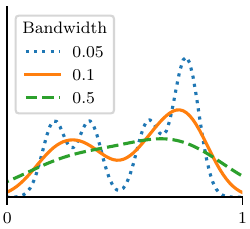}
        \subcaption{A higher bandwidth (wider kernels) smoothes the estimate.}
        \label{fig:overview_kde_bandwidth}
    \end{subfigure}
    \caption{Kernel Density Estimation}
\end{figure}

\paragraph{Density Estimation}
A KDE estimates a distribution by combining
\emph{kernels}, which are centered at each sample and scaled by a
\emph{bandwidth}, which results in a `smoothed line'
through the data points (shown in figure~\ref{fig:overview_kde_illustration}
with gaussian kernels).
The kernel function mainly determines the mathematical
properties of the resulting estimate, such as differentiability,
while the bandwidth determines the trade-off between bias and smoothness
(figure~\ref{fig:overview_kde_bandwidth}).
As evident in the defintion, KDEs require \emph{no} assumption about the
underlying structure of the data, such as normality, and belong to the class
of non-parametric estimators.

Formally, let $x_1, \dots, x_n$ be independent and identically distributed
samples from an unknown probability density \real.
The KDE \est is defined as~\cite[p.~11]{wand_kernel_1995}:
\begin{align}
    \est(x) = \frac{1}{nh} \sum_{i=1}^{n} K \left( \frac{x - x_i}{h}
    \right)
\end{align}
Here, $K$ is the \emph{kernel} (s.t. $\int K(x)~dx = 1$)
and $h$ is the \emph{bandwidth} of the estimator.

\paragraph{Asymptotic Convergence}
KDEs are asymptotically unbiased, i.e.,given sufficient
samples, they can estimate any distribution in the sense that the
\emph{Mean Integrated Square Error} (MISE), which is defined as
\begin{align}
    \mathrm{MISE}(\est)
    = \mathrm{E} \int \left( \est(x) - \real(x)  \right) ^ 2 dx
\end{align}
goes to zero for $n \to\infty$~\cite[p.~23]{wand_kernel_1995}. The Expectation
is taken over all possible training samples of size $n$  drawn from \real.

In particular, for optimal bandwidth, the MISE can be asymptotically
approximated for large sample sizes by
\begin{align}
    \mathrm{MISE}(\est) \approx cn^{-4/5}
    \label{eq:overview_kde_mise}
\end{align}
where the constant factor $c > 0$
depends mainly on $\lvert \real'' \rvert$, i.e. the curvature of the unknown
distribution~\cite[p.~22]{wand_kernel_1995}. It also depends on the
shape of the
kernel function, yet the difference between commonly used
kernel functions is small~\cite[p.~31]{wand_kernel_1995}.

\paragraph{Optimal Estimation}
The bandwidth for optimal asymptotic convergence depends on the unknown true
distribution \real (on $\lvert \real'' \rvert$, to be precise), and it is
subsequently
impossible to pick the optimal
bandwidth, and thus the optimal estimator, a-priori. However, the optimal
bandwidth can be approximated.

Several rules to approximate the optimal bandwidth exist, from simply trying
multiple candidates and selecting the best via cross-validation,
to more sophisticated approaches that try to estimate $\real''$ in order
to compute the optimal bandwidth. While KDEs themselves are non-parametric,
some of these selection rules require assumptions such as normality
and are not suitable for our situation.

In order to remain non-parametric, we use an algorithm known as the
\emph{`Improved Sheather-Jones Algorithm'}~\cite{botev_kernel_2010},
which estimates the optimal bandwidth over a series of iterations and does not
require  normality or other assumptions on the shape of the distribution.
In particular, it is robust (in the sense that it can find a good bandwidth
even for difficult distributions) and can be efficiently computed.

\vspace{0.5cm}

\subsection{Proper Scoring Rules}
\label{ssec:overview_scoring}

As the true distribution is unknown, it is impossible to determine the exact
error of an estimated distribution.
However, it is possible to score the estimate with a test sample.
If the score is computed according to a \emph{proper scoring
    rule}~\cite{gneiting_strictly_2007},
it is closely related to a distance measure, in the sense that maximizing the
score minimizes the distance between the estimated and true distribution.
In particular, \fitnets uses the so called \emph{Quadratic Score} (QS), which
is a proper scoring rule.
The distance measure associated with the QS is the
\emph{Integrated Square Error} (ISE), maximizing the QS
minimizes the ISE.

\paragraph{Propriety}
A scoring rule is \emph{proper}, if it `rewards the best estimate'.
Let $\est$ be an estimated distribution and $x$ a test sample. The score
$\s{\est}{x}$
assigns a real-valued score to observing $x$ for the estimated
distribution.
The score is usually computed with more than a single test value;
let  $\s{\est}{\real}$ denote the expected score w.r.t. test values $x$ drawn
from \real.
The scoring rule is called \emph{proper}, if the correct
estimate achieves the highest score (in expectation), and \emph{strictly
    proper}, if \emph{only}
the correct prediction achieves the highest score.
Formally, a rule is proper, if $\s{\real}{\real} \geq \s{\est}{\real} \
    \forall\ \est,\,\real$ and strictly proper, if this holds with equality if and
only if $\est = \real$~\cite{gneiting_strictly_2007}.

\paragraph{Distance}
Proper scoring rules have an associated distance measure, which is the
difference between the optimal score, i.e. the score of
the true distribution $\real$, and the score of the estimated distribution
$\est$~\cite{gneiting_strictly_2007}:
\begin{align}
    d(\est,\,\real) = \s{\real}{\real} - \s{\est}{\real}
\end{align}

\paragraph{Quadratic Score} The quadratic score is defined
as~\cite{gneiting_strictly_2007}:
\begin{align}
    \qs{\est}{x}
    = \underbrace{2\est(x) \vphantom{\int}}_{reward} -
    \underbrace{\int \est(\omega )^2\,d\omega}_{regularization}
    \label{eqn:overview_qs_def}
\end{align}
and as indicated above, it can be interpreted as combination of a reward for
predicting the correct probability of a test sample and a regularization term,
penalizing overly complex estimates (independent of the test sample).
It's associated distance measure is the integrated square error:
\begin{align}
    d(\est,\,\real)
    = \qs{\real}{\real} - \qs{\est}{\real}
    = \int \left( \est(x) - \real(x)  \right) ^ 2 dx
\end{align}
And in expectation with respect to the sample from which $\est$ is formed, the
MISE is thus related to the distance taken with the expected score of $\est$:
\begin{align}
    \mathrm{MISE}(\est)
    = \qs{\real}{\real} - \mathrm{E} \left[\qs{\est}{\real}\right]
    \label{eq:overview_qs_mise}
\end{align}

\paragraph{Data Plane Scoring}
Given a density estimate \est, computing the score
$\qs{\est}{x}$ for $x$ in a test sample can be done in a single pass over the
test sample, and can be computed in the data plane.
Concretely, computing the mean consists first looking up the reward, i.e.
the estimated probability $\est(x)$, for each incoming value $x$, and
subsequently incrementing counters for both the number of samples and sum of
rewards, in order to allow computation of the mean.
The regularization on the other hand is unrelated to the test sample and can be
computed from the
density estimate alone.

\begin{figure}
    \begin{subfigure}{0.48\linewidth}
        \centering
        \includegraphics[width=\linewidth]{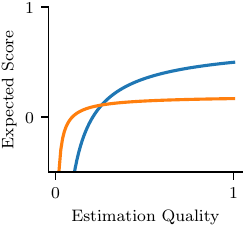}
        \subcaption{Two exemplary scores.}
    \end{subfigure}\hfill
    \begin{subfigure}{0.48\linewidth}
        \centering
        \includegraphics[width=\linewidth]{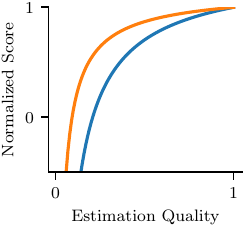}
        \subcaption{The same scores, normalized.}
    \end{subfigure}
    \caption{Normalization makes performance comparable.}
    \label{fig:overview_normalized}
\end{figure}

\vspace{0.5cm}

\subsection{Normalization}
\label{ssec:overview_score_estimation}
Scores for estimated densities of different distributions need to be
\emph{normalized}, otherwise they are not comparable.
However, normalization is not trivial and requires estimation of the optimal
achievable score. \fitnets uses a series of observed scores to compute an
approximate normalization.
Concretely, we formulate this approximation as a constrained
linear optimization problem, which can be efficiently solved.
After normalization, scores can be interpreted as an \emph{estimation
    accuracy}, which is $1$ for the optimal estimate, and lower otherwise.
Additionally, the the solution of the optimization problem can be used to
\emph{predict} the required sample size to reach a certain accuracy, and vice
versa the expected accuracy given a certain sample size, which enables the
sampling rate adaptation of \fitnets.

\paragraph{Comparability requires Normalization}
Proper scoring rules  combine the `predictability' of a
distribution with the quality of the estimate, and thus they need to
be handled carefully when comparing scores of different
distributions~\cite{gneiting_strictly_2007}.
Concretely, a particular score might be close to optimal for one distribution,
but far off for another. Thus, we propose normalizing the score by the optimal
score, such that a score of $1$ always represents the optimal estimate,
regardless of the true distribution (figure~\ref{fig:overview_normalized}):
\begin{align}
    \textrm{NS}(\est) = \frac{\s{\est}{\real}}{\s{\real}{\real}}
\end{align}
Accordingly, we define the \emph{normalized quadratic score} as
$NQS(\est) = \qs{\est}{\real} / \qs{\real}{\real}$.
We note that a normalized  (strictly) proper score is still a (strictly) proper
score, propriety is conserved under affine
transformations~\cite{gneiting_strictly_2007}.

Normalization requires the optimal score $\s{\real}{\real}$, which cannot be
computed exactly without the (unknown) true distribution and can only be
approximated.

\paragraph{Approximate Normalization}
We can approximate the optimal score  for KDEs scored using the quadratic score.
Above, we have shown that the expected distance associated with the quadratic
score is the MISE~(equation~\ref{eq:overview_qs_mise}), which can be
asymptotically approximated for KDEs~(equation~\ref{eq:overview_kde_mise}):
\begin{align}
    \qs{\real}{\real} - \mathrm{E} \left[\qs{\est}{\real}\right]
    \approx c n^{-\sfrac{4}{5}}
    \label{eqn:ls_equation}
\end{align}
The unknown quantities in equation~\ref{eqn:ls_equation} are  $c$ and
\qs{\real}{\real}, which both rely on the unknown true distribution.
The remaining terms, i.e. the training sample size and expected score for this
size, can be empirically measured and subsequently,
equation~\ref{eqn:ls_equation} can be solved for an approximate optimal score.

\paragraph{Constrained Linear Optimization}
An approximate optimal score can be computed by solving a linear
optimization problem.
Concretely, let $\vec{N}$ be a vector of distinct samples sizes, and let
$\vec{S}$ be a vector of mean quadratic scores, s.t.
$\vec{S}_i$ is the mean score of distributions estimated from
samples of size $\vec{N}_i$ from \real.
Then, we estimate the optimal score by solving the following linear
optimization problem:
\begin{alignat}{2}
     & \!\min_{\mathrm{QS}_{opt},\, c} & \quad &
    \lVert\, \vec{1} \cdot \mathrm{QS}_{opt} - \vec{S} - c \vec{N}^{-\sfrac{4}{5}}
    \,\rVert
    \label{eq:overview_opt}
    \\
     & \text{subject to}               &       & \begin{bmatrix}
                                                     \mathrm{QS}_{opt} \\ c
                                                 \end{bmatrix} \geq \begin{bmatrix}
                                                                        \mathrm{QS}_{max} \\ 0
                                                                    \end{bmatrix}
\end{alignat}
where $\mathrm{QS}_{opt} = \qs{\real}{\real}$ and $\mathrm{QS}_{max}$
is the highest score observed for all estimates for the distribution.

Regarding the constraints, the constant $c$ is by definition positive; and as
the quadratic score is proper, the optimal score must be as least as large as
any observed score.

\paragraph{Prediction}
With the solution of the optimization problem we can both predict the
normalized score for a given sampling size, as well as the expected sample size
required to reach a score.
Concretely, the predicted normalized quadratic score for a sample size $n$ is:
\begin{align}
    \mathrm{QS}_{pred}(n) = 1 - \frac{c}{\mathrm{QS}_{opt}}\ n^{-\sfrac{4}{5}}
\end{align}
This function is invertible, and the inverse predicts the required sample size
for a given score.

\paragraph{Practical Improvements}
Repeatedly drawing samples, estimating distributions, and scoring them, in
order to compute the mean score for multiple sample sizes, is impractical in
practice. By using sub-sampling, the required data points for normalization can
be acquired in parallel to the density estimate which is supposed to be
normalized.

Let $S$ be a sample of size $n$, and $\est$ be the KDE computed from $S$. In
order to report \est to the operator, we need a normalized score for
$\est$. We employ the following strategy.
First, select a number of range of sample sizes $n_i < n$ and draw respectively
sized sub-samples $S_i$ from $S$. In parallel to the
estimation of \est, estimate additional densities ${\est}_i$.
Second, score \est and all additional estimates in parallel, which results in
the required data points to solve the optimization problem and normalize the
score for \est without additional steps.

\paragraph{Generalization}
Above, we use the established theory of KDEs combined with quadratic scores to
form a linear optimization problem. However, we argue that this approach can
also be generalized  to other kinds of estimators
and scoring functions under the assumption that the expected distance of an
estimate decreases with increasing sample size, albeit with unknown rate.
This results in a nonlinear optimization problem similar to
\ref{eq:overview_opt}, which requires more resources to solve, yet demonstrates
that the feasibility of our estimation is not limited to KDEs and quadratic
scoring rules (Appendix~\ref{app:normalization}).

%% file: section_design.tex
\begin{table}
    \centering
    \resizebox{\linewidth}{!}{%
        \begin{tabular}{c  l  l  l}
            \textbf{Task} & \textbf{Location}          & \textbf{Constraints}         &
            \textbf{Feature}                                                            \\
            \midrule
            1             & \texttt{Switch 1}          & \texttt{src(42.0.0.0/8)}     &
            \texttt{burst\_size}                                                        \\
            \cmidrule{1-1}\cmidrule{4-4}
            2             &                            &                              &
            \texttt{burst\_duration}                                                    \\
            \cmidrule{1-1}\cmidrule{3-4}
            3             &                            & \texttt{src(43.0.0.0/8)}     &
            \texttt{queue\_time}                                                        \\
            \midrule
            4             & \texttt{Switch 2}          & \texttt{src(13.37.0.0/16) \&
            proto(TCP)}   & \texttt{packet\_size}                                       \\
            \midrule
            \vdots        & \multicolumn{3}{c}{\vdots}
        \end{tabular}
    }
    \captionof{figure}{Monitoring tasks define \emph{where} to monitor
        \emph{which feature} under \emph{which constraints} on the traffic.
        \vspace{0.5cm}
    }
    \label{fig:design_tasks}
\end{table}

\section{System Design}
\label{sec:implementation}

\fitnets consists of four main components, (i) (adjustable) sampling; (ii) density
estimation; (iii) evaluating the estimated densities; and (iv) optimizing the
sampling rate for each task.

In this section, we explain the concrete design of \fitnets from the top down,
first we define the inputs \fitnets requires from the operator, and what it
returns (section~\ref{ssec:design_interface}); next, we explain estimation,
optimization, and adaptation, in the control plane
(section~\ref{ssec:design_cp}); and finally,  we analyze the data plane
processing pipeline (section~\ref{ssec:design_dp}).

\subsection{Interface}
\label{ssec:design_interface}

As introduced in section~\ref{sec:motivation}, operators configure \fitnets with a
list of monitoring tasks and an adaptation objective. In return, \fitnets provides
a stream of estimated probability distributions, along their
accuracy,
for each monitoring task.

\paragraph{Input: Monitoring Tasks}
Monitoring tasks are specified by (i) the \emph{location}, i.e. the
programmable switch, where traffic is monitored; (ii) \emph{constraints},
i.e.,which
traffic is monitored;
and (iii) which \emph{features} are monitored.
For each location, multiple constraints can be specified, and similarly,
for each constraint, multiple features can be monitored.
Figure~\ref{fig:design_tasks} shows a set of exemplary tasks.
\fitnets supports a set of commonly used constraints and features
, which we discuss in the following.
However, we want to note that in general, \fitnets is compatible with any
constraint that can be matched on programmable network devices, as well as any
feature that can be extracted.

For \emph{constraints}, we have implemented the five tuple of source address,
destination address, protocol, source port and destination port,
either individually or in combination, as they are the most common way to
distinguish network traffic.

For \emph{features}, we have implemented a selection of simple features that
can be directly extracted from a single packet, e.g. packet size, as well as
more complex features that require state between packets, such as inter-arrival
time of packets and the duration of flowlets (short burst of packets, separated
by a gap of inactivity).
We focus on traffic bursts as they are both valuable for network monitoring,
and only need to stay in memory for a short time.
Consequently the number of concurrently active flowlets for common timeouts
such as
$100-500ms$ is typically well below $100k$, which is feasible on programmable
network devices (e.g.~\cite{alizadeh_conga:_2014}).
Features that require state for \emph{flows} are a greater challenge. As flows
can be active for minutes, millions of them are typically active at the same
time. Solving this challenge is out of scope for this paper, yet we argue that
it is orthogonal to \fitnets. Any method developed to efficiently handle flow
state on programmable network devices, in software or harware, immediately
extends the range of feature supported by \fitnets.

\paragraph{Input: Adaptation Objective}
Currently, \fitnets supports two common adaptation objectives, either (i)
maximizing the accuracy across all tasks for fixed resources; or (ii)
minimizing the resources for a fixed
accuracy. If needed, it can be easily extended to adapt for other operator
defined objectives.

\paragraph{Output: Scored Densities}
\fitnets processes incoming data in steps, e.g. every second. After each step, it
returns a density  estimate for each monitoring task, along with its accuracy.

\begin{figure}
    \centering
    \includegraphics[width=0.85\linewidth]{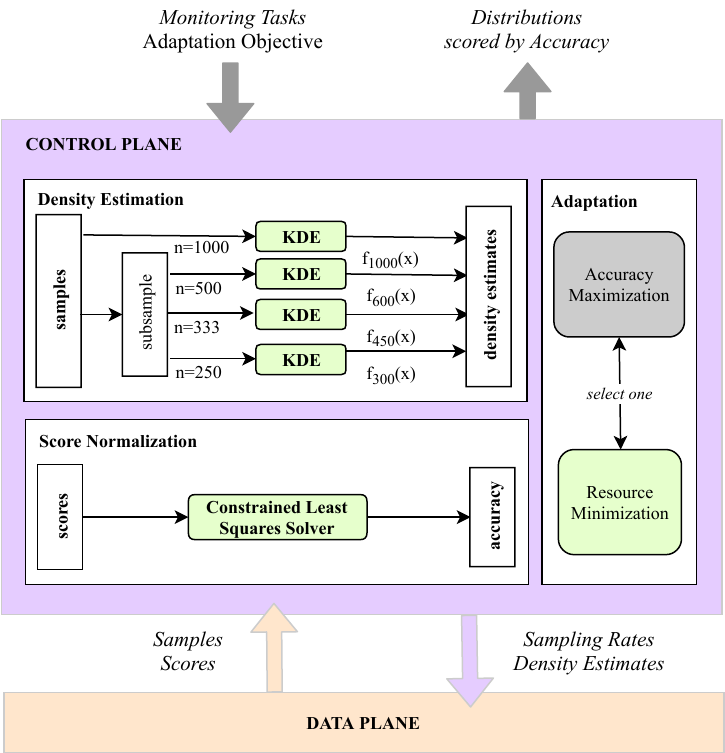}
    \caption{In the control plane, \fitnets estimates density and solves the
        linear optimization problem to compute their accuracy.
        Additionally, it optimizes the sampling rate.}
    \vspace{-0.5cm}
    \label{fig:design_cp}
\end{figure}

\subsection{Control Plane}
\label{ssec:design_cp}

At each computation step, the control plane, (i)
estimates densities; (ii) normalizes scores; and (iii) adapts the sampling rate
per task
(figure~\ref{fig:design_cp}).
Over multiple steps, this forms the following monitoring pipeline:
\begin{description}
    \item[Estimation] At step $i$, density models are estimated from samples
          collected since step $i-1$. The estimated densities are sent back to the
          data plane for scoring.

    \item[Normalization] At step $i$, \fitnets receives the scores for densities
          estimated at step $i-1$.
          \fitnets solves the optimization
          problem required to normalize scores.

    \item[Adaptation] Finally, \fitnets adapts the sampling rate per task,
          optimizing the objective selected by the operator.
\end{description}
In the following, we describe all three steps in further detail, and highlight
how
\fitnets scales through parallelization.

\paragraph{Estimation}
\fitnets needs to estimate multiple densities per task; one density from all the
available data, to be returned to the operator, and additional densities from
subsamples, to enable score normalization to compute the accuracy.

It is possible to re-use a sample to estimate additional
densities by drawing multiple subsamples
(section~\ref{ssec:overview_score_estimation}).
\fitnets employs a simple sub-sampling
scheme of repeatedly splitting the original sample into evenly sized
subsamples, e.g. two subsamples consisting each of a random half of the
original sample, three subsamples  consisting of a third, and so on. This
strategy ensures that for each subsample
size, we use all available data, but do not use data points multiple times,
which could bias the density estimate and thus the mean score.

Regardless of how many densities are estimated, \fitnets only sends a single `mean
density' per task and sample size to the data plane.
The mean of multiple quadratic scores is the mean
of rewards minus the mean of regularization.
Thus, instead of scoring each single model and computing the mean of
scores, \fitnets first computes a single mean density, and sends only this mean
estimate to the data plane for scoring.

\paragraph{Normalization}
From scores received by the data plane, combined with the corresponding sample
sizes from estimation, \fitnets approximates the optimal score per task (as
explained in section~\ref{ssec:overview_score_estimation}).
Concretely, \fitnets uses a constrained least squares solver.
The optimal score is subsequently used to normalize the score of density
estimated from the complete sample, which can subsequently be returned to the
operator.

\paragraph{Adaptation}
\fitnets uses the ability to predict scores to adapt the sampling rate for each
task. After solving the normalization problem, we can estimate the expected
normalized score (i.e. accuracy) for a given sample size $n$
(section~\ref{ssec:overview_score_estimation}). This allows
\fitnets to optimize both adaptation objectives:
\begin{description}
    \item[Resource Minimization] \fitnets predicts the required resources to reach
          the operator-specified accuracy for each task and adapts the sampling rate
          to collect exactly this (minimal) amount of required resources.

    \item[Accuracy Maximization] \fitnets optimizes the \emph{max-min} accuracy
          across all monitoring tasks, i.e. it ensures that the accuracy for the
          least accurate density is maximized, to improve across all tasks.
          For any given accuracy, \fitnets predicts the total resources required.
          Subsequently, finding the specific accuracy that uses all available
          resources, but not more, is a scalar root-finding problem.
\end{description}

\paragraph{Scalability}
The control plane of \fitnets scales to large number of simultaneous tasks, as
most processing steps are independent.
For \emph{each task}, the density estimates for different samples can be
computed in parallel, and parallel to the score normalization.
\emph{Multiple tasks} can not only be parallelized, but also distributed across
multiple machines, as the processing for each task relies only on the samples
and scores for this particular task.
The parallelization of adaptation depends on the objective.For resource
minimization, the individual tasks are independent and consequently, the
adaptation can also be parallelized. Accuracy Maxmimization on the other hand
cannot, as the maximal accuracy depends on the required
resources of any individual task (this affects \emph{only} adaptation,
estimation and normalization can still be parallellized).

\begin{figure*}
    \centering
    \includegraphics[width=\linewidth]{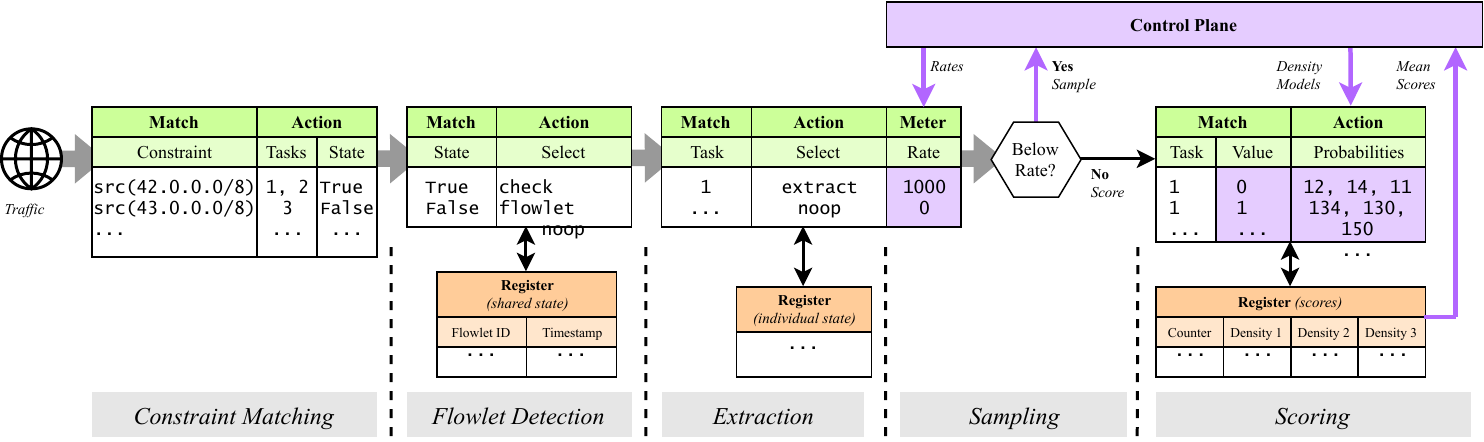}
    \caption{The \fitnets data plane first matches traffic and extract features
        (preprocessing), then either samples the feature or uses it for scoring.
        Multiple features can be processed in parallel, for simplicity, only a
        single one is shown. During preprocessing, flowlet feature require
        state;
        to conserve memory, \fitnets shares flowlet state between features as much
        as
        possible.}
    \label{fig:design_pipeline}
\end{figure*}

\subsection{Data-Plane}
\label{ssec:design_dp}

Conceptually, the data-plane processing consists of three steps, (i)
preprocessing, i.e. matching traffic and extracting features; (ii) sampling;
and (iii) scoring (figure~\ref{fig:design_pipeline}).

In the following, we first summarize the general architecture of programmable
network devices, and next take a closer look at the three
data-plane processing steps.

\paragraph{Architecture}
Programmable switches, programmed using languages such as
P4~\cite{bosshart_p4:_2014}
process packets with a multi-stage pipeline. In each stage, one or more
match-action tables(s) are applied, which can read and modify both packet
headers and additional per-packet metadata.
Additionally, a table can also access objects persisting state between packets,
such as device registers or rate-limiting meters.

A P4 program may be separated into its architecture and configuration.
The architecture defines the control-flow of the program, e.g. table layout or
memory type and size.
Changes to the architecture require re-compilation of the program, and
typically a restart of the switch in order to install the new architecture.
The configuration includes table rules, memory content, and meter rates,
and can be changed at runtime.

The monitoring tasks determine the data-plane architecture of \fitnets.
During setup, the pre- and post-processing steps are programmed and the
required memory for flowlet state and scores is allocated, and meters for each
task are
instantiated. Changes to the monitoring tasks require changes in memory, in
particular for scoring memory, and consequently a recompilation of the program.

\paragraph{Preprocessing}
First, \fitnets extract features from traffic matching
the constraints specified in the monitoring tasks.
For the constraints, \fitnets employs a table matching on the 5-tuple of packet
and  returning the id of all active monitoring tasks.
A monitoring task is active, if its constraint matches.
The table uses ternary matches, which allow flexible queries
such as ranges, longest-prefix-match, covering even complex constraints such as
`UDP packets from the subnet \texttt{42.0.0.0/8} with a destination port
below 1024'.
Additionally, \fitnets checks whether any task requires flowlet state.

If flowlet state is required, \fitnets checks the \emph{flowlet state} of the
current packet using a hash table. The flowlet state can be either (i) the
packet is the start of a new flowlet; (ii) the packet belongs to an active
flowlet; (iii) the packet ends a flowlet (e.g. after a TCP FIN);
or (iv) the state could not be determined because of a hash collision.
\footnote{\fitnets keeps a collision counter such that operators can adjust
    the flowlet state size, as memory cannot be resized at runtime.}
Concretely, \fitnets stores the flow id and the timestamp of the last packet of
the flowlet to both check for collisions and timeouts.

Finally, \fitnets extracts the features of active tasks. For each separate
feature, \fitnets keeps a table that matches on the active tasks.
\footnote{Flowlet feature only match if there has \emph{not} been a collision}
Additionaly, flowlet features might keep further state, e.g. a byte counter for
flowlet sizes.
\fitnets cannot extract every feature after every packet, e.g. the flowlet
duration
can only be returned for the last packet, and excludes un-extracted features
from sampling and scoring.

\paragraph{Sampling}
For each task, \fitnets individually decides whether the feature is sampled or
used for scoring via rate-limiting meters configured from the control plane.
We use `direct meters' attached to the feature extraction tables.
Direct meters keep individual rates for each match of the table and are
automatically executed whenever the table matches. At runtime, the control
plane configures  the rates of the individual matches of the feature extraction
tables, i.e. of the monitoring tasks, as each tasks corresponds to a feature.
Consequently, in the same stage in which features are extracted, the direct
meter output marks the feature for either `sampling' (within the specified
rate) or `scoring' (above the rate).

Finally, \fitnets creates a new packet
containing tuples of \texttt{(task id, feature value)} for all features marked
for sampling, and sends it to the control plane.


\begin{figure*}
    \centering

    \begin{subfigure}[t]{0.2\linewidth}
        \includegraphics[width=\linewidth]{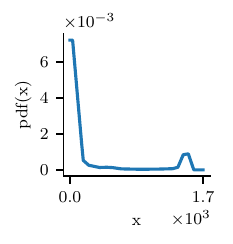}
        \caption{Packet Size [Bytes]}
        \label{fig:eval_dist_size}
    \end{subfigure}%
    \hfill
    \begin{subfigure}[t]{0.2\linewidth}
        \includegraphics[width=\linewidth]{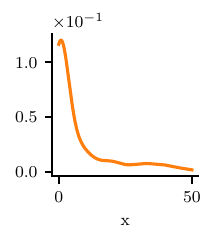}
        \caption{Inter-Arrival Time [ms]}
    \end{subfigure}%
    \hfill
    \begin{subfigure}[t]{0.2\linewidth}
        \includegraphics[width=\linewidth]{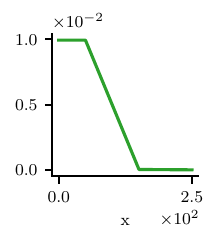}
        \caption{Packets per Flowlet}
    \end{subfigure}%
    \hfill
    \begin{subfigure}[t]{0.2\linewidth}
        \includegraphics[width=\linewidth]{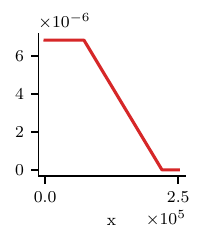}
        \caption{Flowlet Size [Bytes]}
    \end{subfigure}%
    \hfill
    \begin{subfigure}[t]{0.2\linewidth}
        \includegraphics[width=\linewidth]{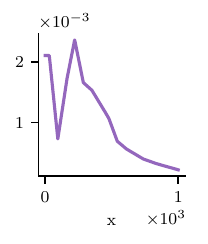}
        \caption{Flowlet Duration [ms]}
        \label{fig:eval_dist_duration}
    \end{subfigure}

    \caption{We evaluate \fitnets with five probability densities
        representative for
        features observed in real network traces. The distributions feature a
        broad range of shapes and scales, ranging from tens of milliseconds, to
        hundred-thousands of bytes.}
    \label{fig:eval_distributions}
\end{figure*}


\paragraph{Scoring}
In the final step of the data-plane pipeline, \fitnets uses unsampled feature
values as test samples to update the estimation scores for the respective tasks.

As described in section~\ref{sec:overview}, \fitnets requires two counters
per  density estimate, one for the number of test samples, and one for the sum
of rewards
(the sum
of $\est(x)$ for all test samples $x$). For
multiple estimates (from multiple training sample sizes), \fitnets can share the
counter for the number of test samples, and thus needs to keep $k+1$ counters
to
score densities estimates from $k$ distinct sample sizes (per task).

In particular, \fitnets implements the lookup of $\est(x)$ using ternary matches.
This allows us to keep the number of table entries
constant, while dynamically adjusting the resolution to the range of \est
(we provide further detail in appendix~\ref{app:lookup}).

%% file: section_evaluation.tex
\vspace*{-5mm}
\section{Evaluation}

In this section we show that \fitnets is accurate for distributions
representative for real network traffic, scales, and is implementable on programmable network devices.

In particular, we analyze:
the flexibility of KDEs to learn arbitrary distributions
(section~\ref{ssec:eval_estimation}), how data-plane scoring ensures reliable
scores (section~\ref{ssec:eval_scoring}),
and that the accuracy of estimates is approximated with an error below $5\%$
(section~\ref{ssec:eval_normalization}).
Furthermore, we evaluate \fitnets's scalability in terms of concurrent estimates
and samples per second for density estimation and in terms of computation time
for the accuracy approximation (section~\ref{ssec:eval_microbenchmarks}).
Finally, we present the two adaptation modes of \fitnets in case studies
(section~\ref{ssec:eval_adaptation}).

\subsection{Methodology}
\paragraph{Ground Truth}
In order to evaluate the performance of \fitnets, in particular the accuracy
of score normalization, we require ground truth distributions.
Thus, we extracted representative distributions for each feature \fitnets
from real-world network traces. We analyzed one hour of the CAIDA backbone
traces~\cite{noauthor_caida_nodate} and collected data for five different
traffic features monitored by \fitnets. For all features that require flowlet state, we have chosen an inter-packet gap of $500ms$, i.e.
after a time gap of $500ms$, a flowlet is considered over.

We  have used KDEs to estimate probability distributions for each feature
(figure~\ref{fig:eval_distributions}) from the observed feature values, which
we use as ground truth for our evaluation.
The five distributions show both a broad range of shapes, e.g. the size
distribution is bimodal (figure~\ref{fig:eval_dist_size}), showing a large peak
for very small packets, and a smaller peak for packets at maximum frame size,
with other sizes mixed in between. The distribution for the observed floatlet
durations shows an even more complex shape
(figure~\ref{fig:eval_dist_duration}).
Furthermore, the distributions als exhibit largely different scales, from tens
of milliseconds for inter-arrival time to hundred-thousands of bytes per burst.
We argue that this diversity is typical for network traffic. While the used
distributions are \emph{representative}, we want to stress that the
distributions represent just a moment in time. In different networks, the same
features are likely distributed differently.

With the distribution, we can in particular determine (i) the expected score of
an estimated density, useful for comparison with empirical scores from test
samples; (ii) the optimal score; (iii) the accuracy or normalized score by
dividing the expected score by the optimal score; and (iv) the integrated
square error (ISE) between a estimate and the true distribution.

\paragraph{Implementation}
We have implemented a multi-core version of the \fitnets control plane in roughly
$1000$ lines of Python code
(using framworks for KDE~\cite{tommy_odland_tommyod/kdepy:_2018} and
optimization~\cite{jones_scipy:_2001}). We process packets using a simulated
switch, i.e. our simulator runs data- and control-plane on a
single machine. In addition, we have implemented the P4 version of the
\fitnets data plane in $\text{P4}_{14}$, $\text{P4}_{16}$ and $\text{P4}_{Tofino}$
(for the Intel
Tofino Switch~\cite{noauthor_barefoot_nodate}) in roughly $500$ lines of P4
code each.


\begin{figure*}
    \begin{minipage}[t]{0.38\linewidth}
        \centering
        \includegraphics[width=\linewidth]{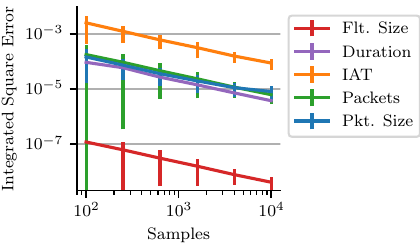}
        \caption{KDEs can approximate all distributions asymptotically, but
            some require a larger sample to be estimated well. }
        \label{fig:eval_density}
    \end{minipage}\hfill%
    \begin{minipage}[t]{0.28\linewidth}
        \centering
        \includegraphics[width=\linewidth]{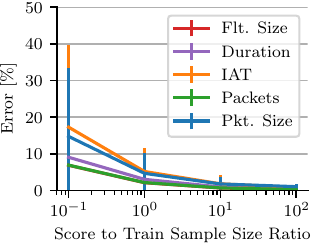}
        \caption{A low relative error in of the empirical score requires a
            large testing sample.}
        \label{fig:eval_scoring}
    \end{minipage}\hfill%
    \begin{minipage}[t]{0.28\linewidth}
        \centering
        \includegraphics[width=\linewidth]{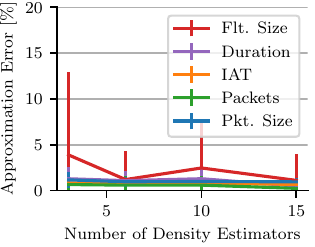}
        \caption{Even with only few estimators, the accuracy is approximated
            with an error below 5\%.}
        \label{fig:eval_normalization}
    \end{minipage}\hfill%
\end{figure*}

\subsection{Density Estimation}
\label{ssec:eval_estimation}

KDEs can approximate any distribution asymptotically for increasing training
sample size, but some distributions require a large sample size than others
(figure~\ref{fig:eval_density}).
In this experiment, we evaluate how KDEs estimate different distributions
by repeatedly drawing samples of increasing size from the each
distribution, computing the KDE, and evaluating the ISE between the estimated
and
true distribution. The figure shows the average ISE along with the standard
deviation across $100$ repetitions per distribution and sample size. As
evident, the error approaches $0$ for all
all distributions, as for all distributions, the ISE steadily decreases
with $n$, while some distributions are harder to estimate than others.
For example, reaching an ISE below $10^{-4}$ requires $100k$ samples for the
inter-arrival time, and less than $1000$ samples for the flowlet duration.

\subsection{Data-Plane Scoring}
\label{ssec:eval_scoring}

Only test samples at about as large as the training sample lead to reliable
($<10\%$ error) scores (figure~\ref{fig:eval_scoring}) across all evaluated
features.
In this experiment, we repeatedly train KDEs
and score each KDE with test samples of varying size,
to compare the empirical scores to the true score of the estimate.

We pick training samples of varying size and determine the test
sample size in relation to the training sample,
ranging from $0.1\times$ to $100\times$ the training
sample size. This reflects typical training situations, e.g. a  data scientist
usually reserves a fraction of the data for testing, e.g. $20\%$;
or in the case of \fitnets, $1\%$ of network traffic might be sampled and used for
estimation, and subsequently $99\%$ for testing.

The figure shows the mean score accuracy for a given ratio of train to test
sample size, along with the standard deviation, over $100$ repetitions for each
distribution, train-, and test sample size. The results for the individual
training sample sizes are comparable, thus we only show the results aggregated
over all training sample sizes.
For test samples smaller than the training sample, the error between the
empirical score and true score varies
strongly, and in cases goes up to $40\%$.
For test sample at least as big as the training sample, the error
is below $10\%$ in most cases, and for even larger test samples, the error
decreases further, approaching $0\%$ for all distributions.

This highlights one of the benefits of benefits of data-plane scoring: \fitnets
does not need to reserve a large fraction of the sampled data to compute
reliable scores. Instead, it profit from typically low sampling rates,
as this ensures that a large number of samples is used for scoring in the
data-plane, resulting in reliable scores.

\medbreak
\subsection{Normalization}
\label{ssec:eval_normalization}

\fitnets approximates  the
estimation accuracy with low error
with only a few additional density estimates
(figure~\ref{fig:eval_normalization}).

\fitnets approximates the estimation accuracy by approximating the optimal
achievable score from additional density estimates, trained with sub-samples of
the training sample (section~\ref{ssec:design_cp}).
In this experiment, we analyze the estimation accuracy, i.e. how well
the returned accuracy matches the true accuracy, depending on the number of
density estimates.

The approximation error across $100$ repetitions is below $15\%$ for $2$
additional estimates, and below $5\%$ for $5$ additional estimates. Our
experiments indicate that using a larger number of
distinct sample sizes with our subsampling strategy does not substantially
improve the performance further, as the the training sample is split into
subsamples of decreasing size, which decreases the quality of the additional
estimators.
Nevertheless, we argue that an approximation error of bewlo 5\% is sufficient
in real-life situations.

\begin{figure*}
    \centering
    \begin{subfigure}{0.49\linewidth}
        \centering
        \includegraphics[width=\linewidth]{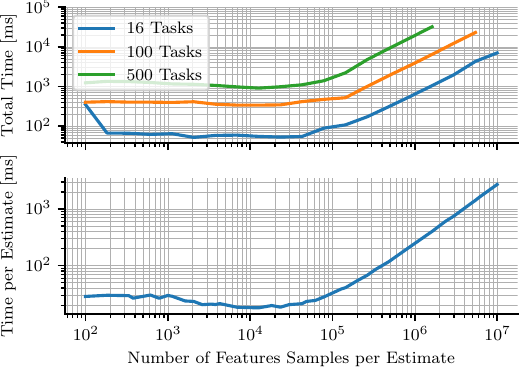}
        \subcaption{Density Estimation}
        \label{fig:eval_benchmark_density}
    \end{subfigure}\hfill%
    \begin{subfigure}{0.49\linewidth}
        \centering
        \includegraphics[width=\linewidth]{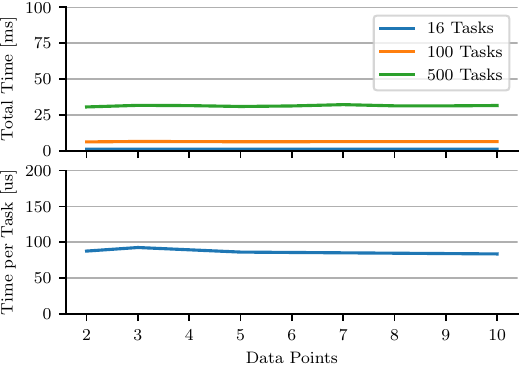}
        \subcaption{Score Normalization}
        \label{fig:eval_benchmark_utility}
    \end{subfigure}\hfill%
    \caption{\fitnets is efficient and parallelizable.}
    \label{fig:eval_benchmarks}
\end{figure*}

\subsection{Benchmarks}
\label{ssec:eval_microbenchmarks}

In this section, we analyze the scalability of \fitnets by measuring the
control-plane execution time. All experiments are repeated $5$ times, and we
show the mean execution time.

First, we analyze the time needed to compute probability densities from
incoming feature samples. \fitnets scales well to large sample sizes, and
\emph{requires roughly $40$ milliseconds to estimate a distribution for up to
    $100k$
    samples} (figure \ref{fig:eval_benchmark_density}). For larger sample sizes,
the time increases linearly, and requires about $1$ second for $3.9M$ samples.
Extending estimation to multiple densities can be efficiently
parallelized, as all estimations are independent. On our (16 core) test machine,
estimating distributions for $16$ densities with $100k$ samples each requires
requires a bit less than $100$ milliseconds (there is some overhead from
initializing worker processes and distributing samples).
In one second, \fitnets can process up to $1.5M$ samples each for $16$ densities
in parallel, and a bit short of $100k$ samples each for 500 densities, which
corresponds to $24M$ and $50M$ samples in total, highlighting that \fitnets can
process millions of samples for up to hundreds of densities in parallel on
a single machine. As discussed in section~\ref{ssec:design_cp}, computation can
be further parallelized by distribution across machines, allowing \fitnets to
scale further.

Next, we analyze the time to solve the optimization problem for normalization,
which \emph{requires less than $10$ milliseconds per optimization}
(figure~\ref{fig:eval_benchmark_utility}).
As described in section~\ref{ssec:design_cp}, solves constrained least
squares problem, which takes less than a millisecond to solve, even for a large
number of inputs. As \fitnets does not require many distinct sample sizes to
compute an accurate normalization (section~\ref{ssec:eval_normalization}), the
computation time of normalization is negligible compared to density estimation requiring less than $50ms$ for $500$ distinct tasks.

\subsection{Case Studies: Adaptive Allocation}
\label{ssec:eval_adaptation}

\begin{figure}
    \begin{minipage}{\linewidth}
        \centering
        \includegraphics[width=0.75\linewidth]{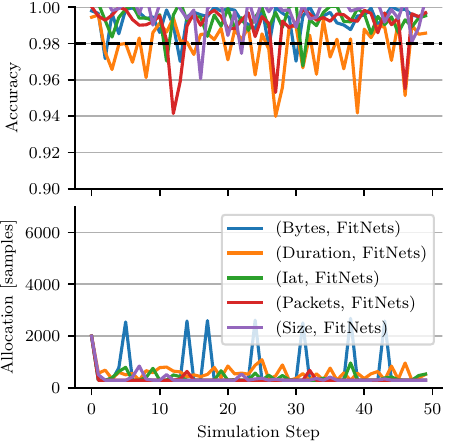}
        \caption{\fitnets can minimize the sampling resources while maintaining
            operator specified accuracy goals (here $0.98$) with low error.}
        \label{fig:eval_min}
    \end{minipage}

    \begin{minipage}{\linewidth}
        \centering
        \includegraphics[width=0.75\linewidth]{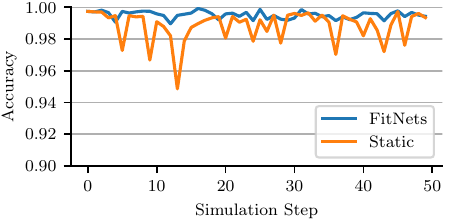}
        \caption{With a fixed bandwidth, \fitnets can achieve a higher accuracy on
            average than static allocation by adaptively adjusting the sampling rate.}
        \label{fig:eval_maxmin}.
    \end{minipage}
\end{figure}

In this section, we demonstrate both adaptation objectives of \fitnets.
The first case study addresses resource minimization. We specify an accuracy
objective of $0.98$ for all traffic features. \fitnets is able to keep the
required accuracy within $\pm 0.02$ in general, and $\pm 0.04$ in the worst
case, while reducing the required samples to a minimum
(figure~\ref{fig:eval_min})

The second case study addresses accuracy maximization. We specify a fixed
bandwidth and compare the performance of \fitnets with a static
allocation. On average, \fitnets achieves higher accuracy with the
same total resources (figure~\ref{fig:eval_maxmin}).

%% file: section_related.tex
\section{Related Work}
\label{sec:related}

\paragraph{In-Network Query Processing}
The processing capabilities of programmable data-planes have given rise to
in-network monitoring. In particular, frameworks such as
Sonata~\cite{gupta_sonata:_2017} and
Marple~\cite{narayana_language-directed_2017} provide an interface similar to
\fitnets: The operator specifies monitoring tasks, in this network queries,
consisting of operators such as \texttt{filter} and \texttt{map}, which are
then executed by the network on all traffic, returning results to the operator.
While the query language itself is expressive, determining the right query for
a given problem can be a complex task in itself, in particular, if the nature
of
the underlying data is unknown. In contrast, \fitnets provides more general
information, estimating distributions for specified traffic statistics.
\fitnets is thus complementing in-network query processing: With the insights
gained from the distributions provided by \fitnets, network operators can design better network queries.

\newpage

\paragraph{Sketches}
Sketches, probabilistic counters, e.g.~\cite{yu_software_2013,liu_one_2016,
  huang_sketchlearn:_2018, yang_elastic_2018}, have seen increasing attention as
they allow to
aggregate data in the network
approximately with provable error bounds.
In particular, it is possible to estimate traffic distributions from the sketch
data structure. However, sketches only apply to \emph{frequency statistics},
i.e.
estimating counts. While this makes them well suited to estimate some traffic
features, such as the flow size distribution, they are not applicable to
others, in particular time-based features, such as the packet inter-arrival
time, which \fitnets offers.
Aside from extracting distributions, the data structure itself is attractive
for
in-network processing, e.g. used by Sonata to reduce the required memory. As
such, they are interesting as a \emph{builing block} for \fitnets, enabling the
extracting of state-intensive features.

\paragraph{Adaptive Stream Processing}
In the context of video streaming, adapting the bandwidth to meet accuracy
objectives has shown to be both feasible and
effective~\cite{rabkin_aggregation_2014}.
In particular, profiling application accuracy for a given bandwidth (both off-
and online) provides a major advance in usability and
performance~\cite{zhang_awstream:_2018}.
The adaptation of \fitnets builds on the same concepts, i.e. estimating the
accuracy and adapting the bandwidth to optimize some objection, on a lower
level. In video streaming, estimating the performance of high-level
applications can be challenging and time-consuming. We have shown that
approximating the accuracy of traffic statistics can be implemented efficiently
in programmable data planes based on proper scoring rules.

%% file: section_conclusion.tex
\newpage
\section{Conclusion}

We introduced \fitnets, a monitoring approach and system for learning accurate
traffic distributions. The key insight behind \fitnets is to combine the control
plane and the data plane through a feedback loop.
In the control plane, \fitnets relies on efficient non-parametric models to learn
distributions of any shape from sampled data. In the data plane, \fitnets tests
the accuracy of learned distributions while dynamically adapting data
collection. By adapting the sampling rate, we show that \fitnets can adapt for a
wide variety of monitoring objectives, including matching required accuracy
requirements or maximizing accuracy given a fixed sampling budget.

We fully implemented \fitnets and show that it can accurately learn representative
for real network traffic traces. We confirmed its practicality by
implementing it on a programmable network device (Intel Tofino).

%% file: appendix.tex
\section{Appendix}

\subsection{Normalization for Arbitrary Estimators and Scores}
\label{app:normalization}

The score normalization presented in
section~\ref{ssec:overview_score_estimation} can be extended to other
estimators and/or other proper scoring rules under the assumption that the
distance associated with the the chosen scoring function can be approximated
by some function
$d(\est,\,\real) \approx c\, n^{-r}$ defined by $c, r > 0$. In other words, we
assume
that the expected distance of the estimate to the true distribution decreases
with increasing sample
size, albeit with unknown rate.
This gives:
\begin{align}
    c n^{-r} \approx
    \s{\real}{\real} - \mathrm{E}\left[\s{\est}{\real}\right]
\end{align}
which, using the same notation as for the linear optimization problem, can be
formulated as the following nonlinear optimization problem:
\begin{alignat}{2}
     & \!\min_{c,\,r,\,S_{opt}} & \quad &
    \lVert S_{opt} - c \vec{N}^{-r} - \vec{S} \rVert        \\
     & \text{subject to}        &       & c, r > 0          \\
     &                          &       & S_{opt} > S_{max}
\end{alignat}

\subsection{Evaluating Probability Densities in the Data-Plane}
\label{app:lookup}
We implement a binning scheme in order to efficiently manage an unknown range
of values for different features.
Using exact matches, the number of entries in the lookup table would equal the
number of distinct feature values. For features such as the flowlet size in
bytes, which can go from $0$ to hundreds of thousands, this is infeasible.

Instead, we use bins, with a width of a power of $2$, which enables us to
fix the number of table rules to a single rule per bin using TCAM matching
(e.g. for bins of size $2^3$, we can set the last three bits in the TCAM match
to 'ignore'). Furthermore, the size of the bins can be adjusted at runtime,
by simply adjusting the matching rules.
\clearpage